# A Simple Mathematical Model of Politics (I)

## Joey Huang

Politics is everywhere. In this paper, I propose a simple model to demonstrate political behavior in human society.

## I. The Matrix $A$

In a society of n persons, the *political status* is represented by an n-by-n matrix $A$. $a_{ij}$ is the *probability* for person $i$ to *listen to* person $j$ when making decisions. If person $i$ is so independent that he only listens to himself, $a_{ii} = 1$ and $a_{ij} = 0$ when $i \neq j$. It's clear that $\sum_{j=1}^{n} a_{ij} = 1$ – the sum of all possibilities is equal to one. Any row of the matrix $A$ is a *probability distribution*.

**Notation**

- $X > Y : x_{ij} > y_{ij}$ for all $i$ and $j$.
- $X > s : x_{ij} > s$ for all $i$ and $j$.

**Definition** An n-by-n matrix $M$ is a *politics matrix* if

- $M\eta = \eta$. $\eta$ is the n-by-1 all-ones vector.
- $M > 0$.

The political status $A$ is *requested* to be a politics matrix – it means $a_{ij} > 0$ for all $i$ and $j$. This property plays an important role in this model. $a_{ij}$ can be very very small – just like it's zero, though. A row of $A$ also plays the role as a *weight function* when the summation of all weights is one.

$\eta$ is a trivial eigenvector of the political status $A$. There's a corresponding left eigenvector.

**Notation** Let $I, J$ be two subsets of $\{1,2,\ldots n\}$. $A_{IJ}$ denotes the submatrix collecting all $a_{ij}$ that $i \in I$ and $j \in J$. $K = I + J$ means $K = I \cup J$ and $I \cap J = \emptyset$. $I'$ denotes the complement of $I$.

**Lemma 1:** Let $u^*$ be a left eigenvector of a politics matrix $A$, $u^*A = u^*$. Then the elements of $u^*$ are either all positive or negative, $u^* > 0$ or $u^* < 0$. In particular, $u^*\eta \neq 0$.

**Proof:** $I = \{i | u_i \geq 0\}, J = \{i | u_i < 0\}$. Assume both $I$ and $J$ are not empty,

$$[u_I^* \ u_J^*] \begin{bmatrix} A_{II} & A_{IJ} \\ A_{JI} & A_{JJ} \end{bmatrix} = [u_I^* \ u_J^*]$$

$$\begin{bmatrix} A_{II} & A_{IJ} \\ A_{JI} & A_{JJ} \end{bmatrix} \begin{bmatrix} \eta_I \\ \eta_J \end{bmatrix} = \begin{bmatrix} \eta_I \\ \eta_J \end{bmatrix}$$

$$u_I^* = u_I^* A_{II} + u_J^* A_{JI}$$

$$u_I^* \eta_I = u_I^* A_{II} \eta_I + u_J^* A_{JI} \eta_I = u_I^* (\eta_I - A_{IJ} \eta_J) + u_J^* A_{JI} \eta_I$$

$$u_I^* A_{IJ} \eta_J = u_J^* A_{JI} \eta_I$$

Because $A > 0$, $A_{IJ} \eta_J > 0$ and $A_{JI} \eta_I > 0$. $u_I^* A_{IJ} \eta_J \geq 0$ and $u_J^* A_{JI} \eta_I < 0$, which is a contradiction. So either $u^* \geq 0$ or $u^* < 0$.

If $u^* \geq 0$, let $P = \{i | u_i > 0\}, O = \{i | u_i = 0\}$. Assume $O$ is not empty,

$$u_P^* A_{PO} \eta_O = (u_P^* A_{PO} + u_O^* A_{OO}) \eta_O = u_O^* \eta_O = 0$$

However, $u_P^* A_{PO} \eta_O > 0$, which is a contradiction.

∎

This all-positive left eigenvector will show up in another way.

## II. The Power

Let's say, people in this society need a leader. $D$ is an n-by-n matrix that $d_{ij}$ is the probability for person $i$ to *support* person $j$. Let's see how people "play politics" to get $D$.

Initially, it's assumed that all the people are *selfish* so they all want themselves to be the leader, 100%. So $D_0 = I$. Note $I$ is the identity matrix and the symbol $I$ is also used as a subset of $\{1,2,\ldots n\}$ in this paper.

Then people start to talk to and convince each other. Up to the political status $A$, each person collects opinions from other people and gets $D_1 = AD_0 = A$.

People keep talking to each other – politics is complicated because of recursive interactions. $D_2 = AD_1 = A^2, D_3 = AD_2 = A^3, \ldots$ So, $D = A^\infty$ if $\lim_{k \to \infty} A^k$ converges.

$A > 0$ plays the key role for the convergence of $A^\infty$ – as a linear operator, $A$ "shrinks" the range of the elements of a vector.

**Lemma 2:** Let $A$ be an n-by-n politics matrix and $v = Au$. $u_M$ and $v_M$ are the maximum elements of vectors $u$ and $v$. $u_m$ and $v_m$ are the minimum elements of vectors $u$ and $v$. If $u_m < u_M$, then $u_m < v_m \leq v_M < u_M$ and there exists a non-negative number $r < 1$ that

$$(v_M - v_m) \leq r(u_M - u_m)$$

**Proof:** Because any row of $A$ is a weight function, any element of $v$ is a weighted mean of all elements of $u$. It's clear that $v_M < u_M$ and $v_m > u_m$.

$A > 0$. Let $\varepsilon \equiv \min_{i,j} a_{ij} > 0$. $1 = \sum_{j=1}^{n} a_{ij} \geq \varepsilon n$. $r \equiv 1 - \varepsilon n \geq 0$

$$a_{ij} = 1 - \sum_{k \neq j} a_{ik} \leq 1 - \varepsilon(n-1)$$

When the largest element of $u$ gets the most possible weight $(1 - \varepsilon(n-1))$ and all other elements get the tiniest weight $\varepsilon$,

$$(1 - \varepsilon(n-1))u_M + \varepsilon(n\bar{u} - u_M) = (1 - \varepsilon n)u_M + \varepsilon n\bar{u} \geq v_M$$

$\bar{u}$ is the mean of all the elements of $u$. The summation of all the elements of $u$ except $u_M$ is $(n\bar{u} - u_M)$.

When the smallest element of $u$ gets the most possible weight $(1 - \varepsilon(n-1))$ and all other elements get the tiniest weight $\varepsilon$,

$$(1 - \varepsilon(n-1))u_m + \varepsilon(n\bar{u} - u_m) = (1 - \varepsilon n)u_m + \varepsilon n\bar{u} \leq v_m$$

So,

$$(1 - \varepsilon n)(u_M - u_m) = r(u_M - u_m) \geq (v_M - v_m)$$

∎

Without $A > 0$, the range is fixed at least but it might not shrink. If the linear operator $A$ is applied to the vector $u$ again and again, the range will shrink to a point. Every column of $D_0 = I$ will converge to $c\eta$. it's easy to get

**Corollary** Let $A$ be an n-by-n politics matrix,

- For any n-by-1 vector $u$, there's a real number $c$ that $\lim_{k \to \infty} A^k u = c\eta$.
- $\lim_{k \to \infty} A^k = A^\infty = \eta\omega^*$. $\omega^*$ is a 1-by-n vector.
- $\lim_{k \to \infty} A^k u = \eta(\omega^* u)$.
- $\eta = A^\infty \eta = \eta(\omega^* \eta)$. So $\omega^* \eta = 1$.
- $\omega^* > 0$.

∎

The unique 1-by-n vector $\omega^*$ is called the *Power* of the politics matrix $A$.

Let $u^*$ be a left eigenvector of $A$, $u^*A = u^*$. $u^*A^\infty = u^* = (u^*\eta)\omega^*$. It's shown in Lemma 1 that $u^*\eta \neq 0$ so $\omega^*$ is a left eigenvector of the politics matrix $A$:

$$\omega^* = \omega^*A$$

Now let's see why this vector is called "power". According to Oxford English Dictionary, power is "the capacity or ability to direct or influence the behavior of others":

$$\omega_j = \sum_{i=1}^{n} \omega_i a_{ij}$$

So, the power of person $j$ is the summation of how much people listen to him ($a_{ij}$) times the power of those people ($\omega_i$) – including himself. It sounds reasonable.

Let's look at this formula in another way:

$$\omega_j = \frac{\sum_{i \neq j} \omega_i a_{ij}}{1 - a_{jj}} = \frac{\sum_{i \neq j} \omega_i a_{ij}}{\sum_{i \neq j} a_{ji}}$$

So, person $j$ gets more power when he listens to himself more and listens to other people less. It's also reasonable.

Because the eigenvalue is known, there's an explicit formula for $\omega^*$. Let $I = \{1\}$ and $J = I'$,

$$[\omega_1 \ \omega_J^*] \begin{bmatrix} a_{11} & A_{1J} \\ A_{J1} & A_{JJ} \end{bmatrix} = [\omega_1 \ \omega_J^*]$$

$$\omega_1 A_{1J} + \omega_J^* A_{JJ} = \omega_J^*$$

$$\omega_J^* = \omega_1 A_{1J}(I - A_{JJ})^{-1}$$

$$\omega^* = \frac{[1 \ A_{1J}(I - A_{JJ})^{-1}]}{1 + A_{1J}(I - A_{JJ})^{-1}\eta_J}$$

$D = A^\infty = \eta\omega^*$. So the possibility for any person to support person $j$ is always $\omega_j$. It's against the common sense. People should have different opinions – even after talking to each other. When all the people have the same probability distribution to decide the leader of this society, it doesn't mean all the people will pick up the same person – but it still doesn't sound real. Something's wrong here – because *selfishness* of human beings is ignored. If all the people are selfish, $D$ stays in $D_0 = I$.

So, under the assumption that all the people are *not* selfish, $\omega_j$ indicates how much person $j$ is *recognized* in this society after people exchange opinions with each other. It also sounds like the *reputation* of person $j$.

Economics assumes selfishness: producers try to maximize their profit and consumers try to maximize their satisfaction. In the next section, *candidates* will be separated from other people in this society and all the people are assumed to be selfish.

## III. The Election

I would like to point out the *rules* of the society are not included in the political status $A$. Democracy or dictatorship or other possible game rules – people play different politics even though the political status stays the same. For example, when the general manager of a firm is determined by the chairman of the board, the main focus of all candidates is this chairman, not other people in this firm.

Now some people are bidding for a position. Let $J$ be the set of all candidates, $I = J'$. It's assumed that all the candidates are selfish so they'll support themselves, 100% ($D_{JJ} = I_{JJ}$). Let $D_{IJ}$ be the matrix that $d_{ij}$ is the probability for person $i$ to support person $j$, $i \in I$ and $j \in J$. People talk to each other to collect different opinions. Let's see what would happen if people were not selfish for the last time. After recursive interactions,

$$\begin{bmatrix} D_{IJ} \\ D_{JJ} \end{bmatrix} = \begin{bmatrix} A_{II} & A_{IJ} \\ A_{JI} & A_{JJ} \end{bmatrix} \begin{bmatrix} D_{IJ} \\ D_{JJ} \end{bmatrix}$$

And

$$\begin{bmatrix} D_{IJ} \\ D_{JJ} \end{bmatrix} = \eta \omega_J^*$$

is a solution of this equation.

Now those candidates are selfish and they always tell people to support themselves when asked for their opinions:

$$D_{JJ} = I_{JJ} \neq [A_{JI} \; A_{JJ}] \begin{bmatrix} D_{IJ} \\ I_{JJ} \end{bmatrix}$$

$$D_{IJ} = [A_{II} \; A_{IJ}] \begin{bmatrix} D_{IJ} \\ I_{JJ} \end{bmatrix}$$

$$D_{IJ} = A_{II} D_{IJ} + A_{IJ}$$

$$D_{IJ} = (I - A_{II})^{-1} A_{IJ}$$

So $D_{IJ}$ is determined by the political status $A$ if $(I - A_{II})$ is invertible.

**Definition** An k-by-k matrix $M$ is a *politics submatrix* if

- $M > 0$.
- $M\eta < 1$. $\eta$ is the k-by-1 all-ones vector.

It's clear that any k-by-k submatrix (k<n) of a political matrix $A$ is a politics submatrix. Person $i$ always listens to person $j$ a little bit – so the sum of the possibilities to listen to people except person $j$ is smaller than one. If $a_{ij}$ can be zero, it's not true.

**Lemma 3** If $M$ is a politics submatrix, then $I - M$ is invertible and

$$(I - M)^{-1} = \sum_{i=0}^{\infty} M^i > 0$$

**Proof** $M\eta < 1$. Let $M\eta \leq r\eta < 1, 1 > r > 0$.

$$M^2\eta = M \cdot M\eta \leq M \cdot r\eta \leq r^2\eta$$

So $M^i\eta \leq r^i\eta$ for $i \geq 0$.

$M > 0$, it's clear that $M^i > 0$ for $i \geq 0$. So $M^i \leq r^i$. The sequence $S_l \equiv \sum_{i=0}^{l} M^i$ is an increasing sequence and $S_l \leq \sum_{i=0}^{l} r^i < 1/(1-r)$. So $S_l$ converges to $S_\infty > 0$ and clearly $S_\infty = (I - M)^{-1}$.

∎

The rank of $I - A$ is smaller than $n$ because $(I - A)\eta = 0$. Assume there's another nonzero vector $\zeta$ such that $(I - A)\zeta = 0$ and $(\eta, \zeta)$ is linearly independent. Without loss generality, let $\zeta_1 = 0$. $J = \{2, \ldots n\}$ and $(I - A_{JJ})\zeta_J = 0$, contradict to Lemma 3 that $I - A_{JJ}$ is invertible. So $rank(I - A) = n - 1$ and $\eta$ and $\omega^*$ play the roles as *the* right and left eigenvectors of the politics status $A$ (for the eigenvalue 1).

**Definition** Let $A$ be an n-by-n politics matrix. $I, J$ are two *disjoint* subsets of $\{1,2,\ldots n\}$.

$$D_{IJ} \equiv (I - A_{II})^{-1} A_{IJ} = -\bar{A}_{II}^{-1} \bar{A}_{IJ}$$

$$\bar{A} \equiv A - I$$

Note $I$ is not requested to be the complement of $J$ in this definition.

It's easy to verify that when $I$ is the complement of $J$,

$$D_{IJ}\eta_J = (I - A_{II})^{-1}A_{IJ}\eta_J = (I - A_{II})^{-1}(\eta_I - A_{II}\eta_I) = \eta_I$$

So every row of $D_{IJ}$ is a probability distribution as expected. The summation of possibilities for person $i$ to support all the candidates is always one.

People play politics to increase the possibilities for other people to support themselves – especially those people whose votes count for the position they bid.

Note that $D_{IJ}$ stays the same when $\bar{A}$ is multiplied by a positive diagonal matrix $\Lambda$. When $\tilde{A} = \Lambda \bar{A}$, $\tilde{A}_{II} = \Lambda_I \bar{A}_{II}$ and $\tilde{A}_{IJ} = \Lambda_I \bar{A}_{IJ}$ so $\tilde{D}_{IJ} = -(\Lambda_I \bar{A}_{II})^{-1}\Lambda_I \bar{A}_{IJ} = D_{IJ}$. In particular, when one row of $\bar{A}$ is multiplied by a positive number, $D_{IJ}$ stays the same. $D_{IJ}$ doesn't collect the behavior of how people support those candidates totally. Let me show it in two simple examples when there are only 4 persons in a society: $I = \{1,2\}, J = \{3,4\}$. Even though all elements of $A$ are requested to be positive, some elements are zero in these two cases so it's easier to handle.

Case 1:

$$A_{II} = \begin{bmatrix} 0.5 & 0.0 \\ 0.5 & 0.0 \end{bmatrix}, A_{IJ} = \begin{bmatrix} 0.4 & 0.1 \\ 0.4 & 0.1 \end{bmatrix}, D_{IJ} = \begin{bmatrix} 0.8 & 0.2 \\ 0.8 & 0.2 \end{bmatrix}$$

Case 2:

$$A_{II} = \begin{bmatrix} 0.5 & 0.0 \\ 0.5\varepsilon & 1-\varepsilon \end{bmatrix}, A_{IJ} = \begin{bmatrix} 0.4 & 0.1 \\ 0.4\varepsilon & 0.1\varepsilon \end{bmatrix}, D_{IJ} = \begin{bmatrix} 0.8 & 0.2 \\ 0.8 & 0.2 \end{bmatrix}$$

Let's calculate the probability for person 4 to get support from both person 1 and person 2. In both cases person 1 doesn't listen to person 2 at all and the probability for him to support person 4 is 20% - then it's up to person 2.

In case 1, there're two possibilities for person 2 to support person 4: listens to person 1 (50%) who decides to support person 4 or listens to person 4 (10%). So in total the probability is 12%=20%*(50%+10%).

In case 2, person 2 is a very independent person that he "almost" listens to himself only. When the small $\varepsilon$ is ignored, he can only support person 4 in one way – listens to himself and decides to support person 4 (20%). So the probability is only 4%=20%*20%.

But in both cases $D_{IJ}$ stays the same. In case 1, 50% person 2 listens to person 1 so it's much more possible for both of them to support the same person. In case 2 they are almost like two independent persons.

To "trace" the behavior of how people support candidates in $J$, every number in $A_{II}$ and $A_{IJ}$ counts ($I = J'$). $D_{IJ}$ shows the probability distribution for every individual in $I$ to support the candidates but the "correlation" is ignored.

## IV. The Families

Let me highlight all important properties derived when all elements of the politics status $A$ are requested to be positive:

- $rank(I - A) = n - 1$.
- $\eta$ and $\omega^*$ are *the* right and left eigenvectors of the politics status $A$.
- $A^\infty = \eta\omega^*$.
- For any submatrix $A_{II}$ of $A$ that $A_{II} \neq A$, $I - A_{II}$ is invertible so $D_{IJ} \equiv (I - A_{II})^{-1}A_{IJ}$ is well-defined.

**Definition** An n-by-n matrix $M$ is a *dominated politics matrix* if

- $M\eta = \eta$. $\eta$ is the n-by-1 all-ones vector.
- $M \geq 0$.

Unlike a politics matrix, the elements of a dominated politics matrix can be 0. The politics status $A$ can be expressed as summation of a dominated politics matrix $\hat{A}$ and a "tiny" term:

$$A = \hat{A} + \varepsilon B, \hat{A}\eta = \eta, B\eta = 0$$

The matrix $B$ takes care of those very small $a_{ij}$. If $\hat{a}_{ij} = 0$, $b_{ij} > 0$. Because $B\eta = 0$, $b_{ij}$ can be negative – when $\hat{a}_{ij}$ is positive.

The power can be expressed as

$$\omega^* = \hat{\omega}^* + \varepsilon\sigma^* + O(\varepsilon^2)$$

$$\hat{\omega}^*\hat{A} = \hat{\omega}^*, \sigma^*\hat{A} + \hat{\omega}^*B = \sigma^*$$

$$\hat{\omega}^*\eta = 1, \sigma^*\eta = 0$$

$\hat{\omega}^*$ is called the *dominated power*.

All the elements of $\omega^*$ are positive because all the elements of $A$ are requested to be positive. Like $\hat{A}$, the elements of $\hat{\omega}^*$ can be 0.

Unlike $I - A_{II}$, $I - \hat{A}_{II}$ can be singular.

**Definition** Let $\hat{A}$ be an n-by-n dominated politics matrix. $F$ is a subset of $\{1,2,\ldots n\}$. $F$ is called a *family* corresponding to the matrix $\hat{A}$ if $\hat{A}_{FF'} = 0$. Or equivalently, $\hat{A}_{FF}$ is also a dominated politics matrix: $\hat{A}_{FF}\eta_F = \eta_F$.

So, people in a family "almost" don't listen to people out of this family. It's like a society inside a society.

$\{1,2,\ldots n\}$ and $\emptyset$ are two trivial families of a dominated politics matrix and they are the *only* two families of a politics matrix. When $A = \hat{A} + \varepsilon B$, there are some nontrivial families possibly based on $\hat{A}$ and the tiny term $\varepsilon B$ can make sure all the "good" properties are still valid – especially $I - A_{II}$ is always invertible. When $I - \hat{A}_{II}$ is singular, the tiny extra term $-\varepsilon B_{II}$ makes $I - A_{II}$ invertible so $D_{IJ}$ is still well-defined. $\hat{D}_{IJ}$ can be defined by $\hat{D}_{IJ} = \lim_{\varepsilon \to 0} D_{IJ}$. It's just like in the era of Newton and Leibniz, $dy$ and $dx$ were "invented" and $dy/dx$ was the derivative.

Note the definition of families is based on the sign of the matrix $\hat{A}$ (1 or 0) only.

**Lemma 4** Let $\hat{A}$ be an n-by-n dominated politics matrix and $I$ and $J$ are two families of $\hat{A}$. Then both $I \cap J$ and $I \cup J$ are families of $\hat{A}$.

**Proof** $\hat{A}_{II'} = 0$, $\hat{A}_{JJ'} = 0$. Let $r \in I \cap J$ and $s \in (I \cap J)' = I' \cup J'$. Either $s \in I'$ or $s \in J'$, $r$ is always a member of both $I$ and $J$ so $\hat{A}_{rs} = 0$. $\hat{A}_{I\cap J,(I\cap J)'} = 0$ – it means $I \cap J$ is a family. Similarly, it's easy to conclude $I \cup J$ is a family as well.

∎

So, the collection of all families is a *topology* on the finite set $\{1,2,\ldots n\}$. Let me remind the definition of a topology:

**Definition** $\tau$ is a collection of subsets of $X$. $\tau$ is called a topology on $X$ if it satisfies the following axioms:

- $X$ and $\emptyset$ belong to $\tau$.
- Any arbitrary union of members of $\tau$ still belongs to $\tau$.
- The intersection of any finite number of members of $\tau$ still belongs to $\tau$.

The definition of a topology is very abstract but it seems to come out naturally based on a natural definition of families in a society. In 1966, H. Sharp [1] and V. Krishnamurthy [2] showed a topology on a finite set can be represented by a Boolean matrix. Topology was not invented to handle finite sets so the research didn't get a lot of attention – it's like an abstract game played by few mathematicians. Surprisingly, the number $T_n$ of possible topologies on $\{1,2,\ldots n\}$ is still an open problem [3].

**Lemma 5** Let $\hat{A}$ be an n-by-n dominated politics matrix. $I$ is a non-empty subset of $\{1,2,\ldots n\}$. If $I - \hat{A}_{II}$ is singular and $u_I^*$ is a non-zero left eigenvector that $u_I^*(I - \hat{A}_{II}) = 0$, then $G = \{i | i \in I, \ u_i > 0\}$ is a family. Similarly, $\{i | i \in I, \ u_i < 0\}$ is also a family. In particular, *there's always a non-empty family inside I if $I - \hat{A}_{II}$ is singular.*

**Proof** If $u_I^* > 0$, $0 = u_I^*(I - \hat{A}_{II})\eta_I = u_I^*\hat{A}_{II'}\eta_{I'}$, $\hat{A}_{II'}\eta_{I'} = 0$. So $\hat{A}_{II'} = 0$ and $G = I$ is a family.

If some components of $u_I^*$ are not positive, $I = G + H, u_G^* > 0, u_H^* \leq 0$. If $G = \emptyset$, of course it's a family. If not,

$$[u_G^* \ u_H^*] = [u_G^* \ u_H^*]\begin{bmatrix}\hat{A}_{GG} & \hat{A}_{GH} \\ \hat{A}_{HG} & \hat{A}_{HH}\end{bmatrix}$$

$$u_G^*\eta_G = (u_G^*\hat{A}_{GG} + u_H^*\hat{A}_{HG})\eta_G$$

$$u_H^*\hat{A}_{HG}\eta_G = u_G^*(\eta_G - \hat{A}_{GG}\eta_G) = u_G^*(\hat{A}_{GH}\eta_H + \hat{A}_{GI'}\eta_{I'})$$

Because $u_H^*\hat{A}_{HG}\eta_G \leq 0$ & $u_G^* > 0$, $\hat{A}_{GH}\eta_H + \hat{A}_{GI'}\eta_{I'} = 0$. So $\hat{A}_{GG'} = \hat{A}_{G,H+I'} = 0$. $G$ is a family.

$-u_I^*$ is also an non-zero left eigenvector so $\{i | i \in I, \ u_i < 0\}$ is also a family.

∎

**Lemma 6** Let $\hat{A}$ be an n-by-n dominated politics matrix. $I$ is a non-empty subset of $\{1,2,\ldots n\}$. Then $I - \hat{A}_{II}$ is singular *if and only if* there's a non-empty family $F \subset I$.

**Proof** If there's a non-empty family $F \subset I$ and $I = F + G$, $\hat{A}_{FG} = 0$ and $\hat{A}_{FF}\eta_F = \eta_F$. So $I_F - \hat{A}_{FF}$ is singular. Let $u_F^*$ be a non-zero left eigenvector of $I_F - \hat{A}_{FF}$, $u_F^*(I_F - \hat{A}_{FF}) = 0$. Then

$$[u_F^* \ 0](I - \hat{A}_{II}) = [u_F^* \ 0]\begin{bmatrix}I_F - \hat{A}_{FF} & 0 \\ -\hat{A}_{GF} & I_G - \hat{A}_{GG}\end{bmatrix} = 0$$

So $I - \hat{A}_{II}$ is singular.

Only if: It's part of Lemma 5.

∎

Let me describe Lemma 6 in one sentence: *Families create Singularities.*

**Definition** Let $\hat{A}$ be an n-by-n dominated politics matrix. A non-empty family $U$ corresponding to $\hat{A}$ is called an *upper-class family* if $U$ and $\emptyset$ are the only two families inside $U$. In other words, there's no "subfamily" inside an upper-class family.

Let $U_1$ and $U_2$ be two upper-class families and $U_1 \cap U_2 \neq \emptyset$. Because $U_1 \cap U_2$ is also a family and it's inside both $U_1$ and $U_2$, $U_1 = U_1 \cap U_2 = U_2$. It's clear that all upper-class families are *disjoint*.

Because this topology is defined on a finite set, it means any non-empty family contains at least one upper-class family inside – it can be itself. So Lemma 6 is still valid when "non-empty family" is replaced by "upper-class family".

**Lemma 7** Let $\hat{A}$ be an n-by-n dominated politics matrix. $U$ is an upper-class family corresponding to $\hat{A}$ and $u_U^*$ is an non-zero left eigenvector that $u_U^*(I - \hat{A}_{UU}) = 0$.

- Either $u_U^* > 0$ or $u_U^* < 0$.
- $rank(I - \hat{A}_{UU}) = k - 1$. $k$ is the number of elements in $U$.

**Proof** If one component of $u_U^*$ is positive, by Lemma 5 $\{i | i \in U, u_i > 0\}$ is a family and it's not empty. Because $U$ is an upper-class family, this set is $U$ and $u_U^* > 0$. Similarly, if one component of $u_U^*$ is negative, $u_U^* < 0$.

Let $u_U^* > 0$. If $rank(I - \hat{A}_{UU}) < k - 1$, there's another non-zero left eigenvector $\tilde{u}_U^* > 0$ and $(u_U^*, \tilde{u}_U^*)$ is linearly independent. It's easy to find a number $c$ that one component of $u_U^* + c\tilde{u}_U^*$ is 0 when it's still a non-zero left eigenvector of $I - \hat{A}_{UU}$, which is a contradiction.

∎

The next Lemma shows people who don't belong to any upper-class family won't get any dominated power. So the name "upper-class" has the right meaning.

**Lemma 8** Let $\hat{A}$ be an n-by-n dominated politics matrix. $\hat{\omega}^*$ is the dominated power that $\hat{\omega}^*\hat{A} = \hat{\omega}^*$, $\hat{\omega}^* \geq 0$. $U$ is the union of all upper-class families corresponding to $\hat{A}$. Then $\hat{\omega}_{U'}^* = 0$.

**Proof** $U$ is a family. $\hat{A}_{UU'} = 0$.

$$[\hat{\omega}_U^* \ \hat{\omega}_{U'}^*] \begin{bmatrix} \hat{A}_{UU} & 0 \\ \hat{A}_{U'U} & \hat{A}_{U'U'} \end{bmatrix} = [\hat{\omega}_U^* \ \hat{\omega}_{U'}^*]$$

$\hat{\omega}_{U'}^* \hat{A}_{U'U'} = \hat{\omega}_{U'}^*$. There's no (upper-class) family inside $U'$ so $I - \hat{A}_{U'U'}$ is invertible by Lemma 6. $\hat{\omega}_{U'}^*(I - \hat{A}_{U'U'}) = 0$ so $\hat{\omega}_{U'}^* = 0$.

∎

Lemma 8 shows why people in a society are motivated to form families. In a well-connected society where any two persons "really" listen to each other, there're only two families: ∅ and $\{1,2,\dots n\}$. Let's say, *We Are the World*. This kind of ideal world is hard to achieve because people are selfish. People are willing to get more power than other people.

Let's consider a very simple case: there's only one nontrivial family $F$ in a society. Because it's the only one, $F$ is an upper-class family. Let's see the benefit they get mathematically. The politics status looks like

$$\begin{bmatrix} A_{FF} & \varepsilon B_{FF'} \\ A_{F'F} & A_{F'F'} \end{bmatrix}$$

Now people are bidding for a position. $\{1,2,\dots n\} = I + J$ and $J$ is the set of all candidates. When some people $K$ in the family are candidates, $I = G + H, J = K + L$ and $A_{II}$ $A_{IJ}$ look like

$$A_{II} = \begin{bmatrix} A_{GG} & \varepsilon B_{GH} \\ A_{HG} & A_{HH} \end{bmatrix}, A_{IJ} = \begin{bmatrix} A_{GK} & \varepsilon B_{GL} \\ A_{HK} & A_{HL} \end{bmatrix}$$

Where $F = G + K$ and $F' = H + L$.

The only nontrivial family $F$ is not inside $I$ so $I - \hat{A}_{II}$ is invertible:

$$(I - \hat{A}_{II})^{-1} = \begin{bmatrix} (I_G - \hat{A}_{GG})^{-1} & 0 \\ (I_H - \hat{A}_{HH})^{-1}\hat{A}_{HG}(I_G - \hat{A}_{GG})^{-1} & (I_H - \hat{A}_{HH})^{-1} \end{bmatrix}$$

$$(I - \hat{A}_{II})^{-1} = \begin{bmatrix} (I_G - \hat{A}_{GG})^{-1} & 0 \\ \hat{D}_{HG}(I_G - \hat{A}_{GG})^{-1} & (I_H - \hat{A}_{HH})^{-1} \end{bmatrix}$$

$$\hat{D}_{IJ} = (I - \hat{A}_{II})^{-1}\hat{A}_{IJ} = \begin{bmatrix} \hat{D}_{GK} & 0 \\ \hat{D}_{HG}\hat{D}_{GK} + \hat{D}_{HK} & \hat{D}_{HL} \end{bmatrix}$$

(1)

$$D_{IJ} = \hat{D}_{IJ} + O(\varepsilon)$$

As expected, people in this family almost won't support people out of the family ($0 + O(\varepsilon)$). They support people in this family almost like how they support them "internally" in this family $F = G + K$ ($\hat{D}_{GK} = (I_{GG} - \hat{A}_{GG})^{-1}\hat{A}_{GK}$).

For those people who are not in this family, they support those candidates like all the people in family $F$ were candidates:

$$\hat{D}_{H,G+K+L} = [\hat{D}_{HG}\ \hat{D}_{HK}\ \hat{D}_{HL}]$$

Not really – people in $G$ are not candidates. $\widehat{D}_{HG}$ is not useless – it plays the role as a linear operator to take the internal support inside family $F$ ($\widehat{D}_{GK}$) into consideration ($\widehat{D}_{HG}\widehat{D}_{GK}$). They are not "slaves" of the upper-class family, though. They have their own opinions ($\widehat{D}_{HK}$).

People in the family can act like other people "almost" don't exist. Their internal opinions ($\widehat{D}_{GK}$) are still "valid" – up to a linear operator for people who are not in this family. The benefit of this (upper-class) family is evident.

When no candidate belongs to this family $F$, $F$ is inside $I$ so $I - \hat{A}_{II}$ is singular. The inverse of an almost singular matrix can be calculated by the next lemma (the first two dominated terms). The proof is omitted.

**Lemma 9** Let $M$ be a singular n-by-n matrix. $rank(M) = n - k, 0 < k < n$. Pick up a subset $G$ of $\{1,2,\dots n\}$ with $n - k$ elements that $M_{GG}$ is invertible. Collect $k$ independent right vectors as an n-by-k matrix $V$ and another $k$ independent left vectors as an k-by-n matrix $U^*$ such that

$$MV = 0, U^*M = 0$$

Let $N$ be another n-by-n matrix. $\Omega \equiv U^*NV, W \equiv V\Omega^{-1}U^*$. Then

$$(M + \varepsilon N)^{-1} = W\varepsilon^{-1} + (I - WN)Z(I - NW) + O(\varepsilon)$$

(2)

Where

$$Z = \begin{bmatrix} M_{GG}^{-1} & 0 \\ 0 & 0 \end{bmatrix}$$

∎

$I = F + H$. $A_{II}$ and $A_{IJ}$ look like

$$A_{II} = \begin{bmatrix} A_{FF} & \varepsilon B_{FH} \\ A_{HF} & A_{HH} \end{bmatrix}, A_{IJ} = \begin{bmatrix} \varepsilon B_{FJ} \\ A_{HJ} \end{bmatrix}$$

$$I - A_{II} = \begin{bmatrix} I_F - \hat{A}_{FF} & 0 \\ -\hat{A}_{HF} & I_H - \hat{A}_{HH} \end{bmatrix} - \varepsilon \begin{bmatrix} B_{FF} & B_{FH} \\ B_{HF} & B_{HH} \end{bmatrix}$$

The left eigenvector of $\hat{A}_{II}$ is $[\widehat{\omega}_F^* \ 0]$ – it's shown in Lemma 8 that $\widehat{\omega}_H^* = 0$. The right eigenvector is

$$\begin{bmatrix} \eta_F \\ \widehat{D}_{HF}\eta_F \end{bmatrix}$$

The formula in Lemma 9 is applied to get $(I - A_{II})^{-1}$ and $\widehat{D}_{IJ}$ as well:

$$c_J^* = \widehat{\omega}_F^*(B_{FJ} + B_{FH}\widehat{D}_{HJ}), \bar{c}_J^* = \frac{c_J^*}{c_J^* \eta_J}$$

$$\widehat{D}_{IJ} = \begin{bmatrix} \eta_F \bar{c}_J^* \\ \widehat{D}_{HF} \eta_F \bar{c}_J^* + \widehat{D}_{HJ} \end{bmatrix}$$

(3)

$$D_{IJ} = \widehat{D}_{IJ} + O(\varepsilon)$$

In this case the family doesn't almost ignore other people totally. Let's look at people in $H$ first. Once again, if all the people in the family were candidates, this is how they support $F + J$:

$$\widehat{D}_{H,F+J} = [\widehat{D}_{HF} \; \widehat{D}_{HJ}]$$

People in the family still listen to other people slightly ($\varepsilon B_{FH}, \varepsilon B_{FJ}$) and $B_{FH}$ plays the role as a linear operator to take $\widehat{D}_{HJ}$ into consideration. A *consensus* ($\bar{c}_J^*$) weighted by the power ($\widehat{\omega}_F^*$) of the family is obtained.

All the members in the family almost support the candidates based on the consensus ($\eta_F \bar{c}_J^* + O(\varepsilon)$). Note it doesn't mean they will almost support the same person.

$\widehat{D}_{HJ}$ plays a role in the consensus and then the consensus also plays a role in how other people support the candidates ($\widehat{D}_{HF} \eta_F \bar{c}_J^* + \widehat{D}_{HJ} + O(\varepsilon)$) – up to the vector $\widehat{D}_{HF} \eta_F$.

To get more support, the candidates $J$ need to "maintain" $A_{FJ} = \varepsilon B_{FJ}$ and $A_{HJ} = \widehat{A}_{HJ} + \varepsilon B_{HJ}$. $c_J^* = \widehat{\omega}_F^*(B_{FJ} + B_{FH}\widehat{D}_{HJ}) = \widehat{\omega}_F^*\left(B_{FJ} + B_{FH}(I_H - \widehat{A}_{HH})^{-1}\widehat{A}_{HJ}\right)$. Even though people in the family only listen to those candidates slightly, $B_{FJ}$ is as important as $\widehat{A}_{HJ}$ in this formula. That's the benefit of the family as well. For example, if 0.1% someone in $F$ listens to you and 50% another person in $H$ listens to you, it might be a good deal when 0.1% becomes 1% and 50% becomes 40%.

And it's clear that you need to "flatter" people with more power more. The name "power" has the right meaning.

Other people will just sit and see the benefit this family gets? Probably not.

**Definition** Let $\widehat{A}$ be the n-by-n dominated politics matrix of a society $\{1,2,...n\}$. This society is *disconnected* if it is the union of two disjoint non-empty families. Otherwise, it's a *connected* society.

This definition is identical with the definition of connectivity in a topology space. In this model, an abstract definition in topology becomes so natural and it's so related to the real world.

**Lemma 10** Let $\hat{A}$ be the n-by-n dominated politics matrix of a society $\{1,2,\ldots n\}$. If this society is connected and there are more than one upper-class family, then the union of all upper-class families $U$ cannot be the whole society, $U' \neq \emptyset$. In particular, $\hat{\omega}_i^* = 0$ for some $i \in \{1,2,\ldots n\}$.

**Proof** All upper-class families are disjoint. If $\bigcup_{j=1}^{q} U_j = U = \{1,2,\ldots n\}, q > 1$, the society is the union of two disjoint non-empty family $U_1$ and $\bigcup_{j=2}^{q} U_j$, which is a contradiction. So $U' \neq \emptyset$. $\hat{\omega}_{U'}^* = 0$ by Lemma 8.

∎

If there are at least two upper-class families in a society, some people need to "sacrifice" their power to keep the society connected. But why do they want to sacrifice?

In conclusion, "We Are the World" is probably just a song. A society without any nontrivial family is hard to be achieved. When some people come together and form a family to get some benefit, other people won't be happy and some of them might form other families to get their own benefit as well. The best example is the *Cold War* (1947-1991). A new cold war is coming and it looks like nobody can stop it.

## V. Garden of Eden

A society without any nontrivial family is hard to be achieved. Let's look the other way: how about a society where people almost don't "play politics"?

This is an almost-no-politics garden:

$$\hat{A} = I, A = I + \varepsilon B, \bar{A} = A - I = \varepsilon B$$

$$D_{IJ} = -\bar{A}_{II}^{-1}\bar{A}_{IJ} = -B_{II}^{-1}B_{IJ}$$

People in this garden are all smart and moral. They are smart enough to have their own opinions and they "listen to their hearts" almost 100% ($I$). They still interact with other people to collect some points they might ignore ($\varepsilon B$). When some candidates are bidding for a position, other people exchange ideas lightly and then almost make their own decisions individually ($D_{IJ} = -B_{II}^{-1}B_{IJ}$). Note that $\varepsilon$ doesn't appear in the formula of $D_{IJ}$. It's shown that $D_{IJ}$ stays the same when $\bar{A}$ is multiplied by a diagonal matrix. A positive number – large or small – doesn't change $D_{IJ}$, either.

In this case, every individual is an upper-class family and any subset of $\{1,2,\ldots n\}$ is a family.

This ideal society doesn't really exist. First, many people are not smart enough to have their own opinions. We can collect some smart people to form an artificial society – let's say, it's really like an almost-no-politics garden initially. Later, some people decide to play "real" politics and form an upper-class family $F$. The politics status becomes

$$A = \begin{bmatrix} A_{FF} & \varepsilon B_{FF'} \\ \varepsilon B_{F'F} & I_{F'} + \varepsilon B_{F'F'} \end{bmatrix}$$

$$\bar{A} = \begin{bmatrix} A_{FF} - I_F & \varepsilon B_{FF'} \\ \varepsilon B_{F'F} & \varepsilon B_{F'F'} \end{bmatrix}$$

$D_{IJ}$ stays the same when different rows of $\bar{A}$ are multiplied by different positive numbers. So it can be calculated by

$$\tilde{A} = \begin{bmatrix} A_{FF} - I_F & \varepsilon B_{FF'} \\ B_{F'F} & B_{F'F'} \end{bmatrix}$$

It's just like the case when $F$ is the only non-trivial family of this society. Those single-member upper-class families are "useless" when nobody listens to them "seriously". Equation (1) and (3) are still valid and so is the benefit.

Now people in this family have eaten the "forbidden fruit". There're three options for other people:

- Join the family to share the "forbidden fruit".
- Form their own family to get their own "forbidden fruit".
- Stay the same and sing "Bohemian Rhapsody" – nothing really matters, anyone can see.

An almost-no-politics society is not *stable* so it doesn't exist.

Let's recall the romantic poem written by Alexander Pope:

*Nature and Nature's laws lay hid in night*

*God said, Let Newton be!*

*And all was light*

An unromantic poem can be written in a similar way:

*The idea of playing politics lay hid in night*

*Devil said, Let's form a family!*

*And politics was everywhere*

Politics is unavoidable. It's possible that those people who refuse to play "real" politics are isolated in a society eventually – even though every individual is an upper-class family by definition.

## VI. Some Family Structures

The simplest family structure is

**Father and Sons**

$$\hat{A}_{FF} = \eta \cdot [1\ 0\ 0\ ...\ 0]$$

It's not an upper-class family. There's only one upper-class family inside – the leader himself. A single-person upper-class family works when some people do listen to him "seriously". Of course the leader gets the support from almost all the people in the family when he is a candidate. What will happen if the leader is not a candidate? Let's see it *internally* first. If the family is the whole society, Equation (3) can be applied in this case:

Candidates: $J$. $J' = I = \{1\} + H$.

$$A_{II} = \begin{bmatrix} 1 + \varepsilon b_{11} & \varepsilon B_{1H} \\ \eta_H + \varepsilon B_{H1} & \varepsilon B_{HH} \end{bmatrix}, A_{IJ} = \begin{bmatrix} \varepsilon B_{1J} \\ \varepsilon B_{HJ} \end{bmatrix}$$

$$\bar{c}_J^* = \frac{B_{1J}}{B_{1J}\eta_J}, \hat{D}_{IJ} = \eta_I \bar{c}_J^*$$

(4)

$$D_{IJ} = \hat{D}_{IJ} + O(\varepsilon)$$

So, the leader picks up someone based on how he listens to those candidates slightly ($\varepsilon B_{1J}$), then all other people will almost listen to the leader and support this person as well. That's why the probabilities for them to support those candidates are almost the same ($\eta_I \bar{c}_J^* + O(\varepsilon)$).

When the family is not the whole society, the leader also listens to people out of the family slightly so people in the family don't really get any privilege. The leader can pick up someone who doesn't belong to this family, too. Why should they almost always support this leader when they get no benefit? It's more reasonable that

$$\hat{A}_{FF} = \begin{bmatrix} \hat{a}_{11} & \hat{a}_{12} & ... & \hat{a}_{1k} \\ 1 & & & \\ \vdots & & 0 & \\ 1 & & & \end{bmatrix}, \Sigma_j\, \hat{a}_{1j} = 1$$

All the people in this family almost listen to the leader and the leader listens to all the members up to the vector $\hat{A}_{1F}$. It looks like this: the leader collects opinions from people in the family and picks up a candidate – then the whole family will almost all support this candidate. The leader only listens to people out of the family slightly ($\varepsilon$) – so people in this family get the support they deserve when they are candidates of a position. In this case the family $F$ is an upper-class family and $B_{1J}$ is replaced by $\hat{A}_{1J}$ in Equation (4).

**Family Tree**

This is the standard structure of a family. It's widely used in armies, firms and political parties. Note the political status shows the "real" status of a society and it can be different from the official reporting system. In an army, it should be almost identical with the official structure because the punishment is very severe when someone doesn't follow an order – $\varepsilon B$ should be very very small. In a mafia, all the positions are not "official" so it relies on the political status totally. The "spiritual leader" doesn't really have an ID indicating his "leadership". In a company, the son of the owner probably doesn't need to listen to his manager.

Let's see how this kind of structure works in this simple model. The dominated politics matrix of a family looks like this, for example:

$$\hat{A} = \begin{bmatrix} 1 & 0 & 0 & 0 & 0 & 0 \\ 1 & 0 & 0 & 0 & 0 & 0 \\ 1 & 0 & 0 & 0 & 0 & 0 \\ 0 & 1 & 0 & 0 & 0 & 0 \\ 0 & 1 & 0 & 0 & 0 & 0 \\ 0 & 0 & 1 & 0 & 0 & 0 \end{bmatrix}$$

Person 1 is the big leader so he almost listens to himself only; Person 2 and 3 almost listen to person 1 totally; Person 4 and 5 "report" to person 2; Person 6 reports to person 3.

Again, it's not an upper-class family – the big leader himself is the only upper-class family inside this family. When the big leader is the only candidate of a position in this family, all the people in the family will almost support him. However, that's not the case when more people in this family are also candidates. People are selfish. In this example, when both person 1 and person 2 are candidates, person 4 and 5 prefer to support person 2. Person 3 and person 6 still prefer person 1, though. Anyway, when the big leader is one of the candidates, the preferences of all the members in this family are well-defined based on the tree structure.

The big leader needs to be "careful". The best example is Mikhail Gorbachev.

When the big leader is not a candidate, the preference of the big leader decides the preferences of people who still "listen to him". Again, let's see what it is internally when the family is the whole society. People in the family are grouped in $q + 3$ subsets: $\{1\}, G_0, G_1, G_2 \ldots G_q, J$. There are $q$ candidates in $J$ and people in $G_k, 0 < k \leq q$ prefer the $k$th candidates based on the tree structure. People in $G_0$ still "listen to" person 1. $I = \{1\} + G_0 + G_1 + G_2 + \cdots G_q$. $\hat{A}_{II}$ looks like

$$\hat{A}_{II} = \begin{bmatrix} 1 & 0 & & & 0 \\ \hat{A}_{G_0 1} & \hat{A}_{G_0 G_0} & & & \\ & & \hat{A}_{G_1 G_1} & & 0 \\ & 0 & & \ddots & \\ & & 0 & & \hat{A}_{G_q G_q} \end{bmatrix}$$

$\hat{A}_{IJ}$ looks like

$$\hat{A}_{IJ} = \begin{bmatrix} 0 & 0 & 0 \\ 0 & 0 & 0 \\ \hat{A}_{G_1 j_1} & & \\ & \ddots & \\ & & \hat{A}_{G_q j_q} \end{bmatrix}$$

$G = G_0 + G_1 + G_2 + \cdots G_q$. Equation (3) is applied again:

$$\hat{D}_{GJ} = \begin{bmatrix} 0 & 0 & 0 \\ \eta_{G_1} & & \\ & \ddots & \\ & & \eta_{G_q} \end{bmatrix}$$

$$c_J^* = B_{1J} + \begin{bmatrix} B_{1G_1} \eta_{G_1} & B_{1G_2} \eta_{G_2} & \cdots & B_{1G_q} \eta_{G_q} \end{bmatrix}, \bar{c}_J^* = \frac{c_J^*}{c_J^* \eta_J}$$

(5)

This "consensus" is for the big leader and all the people in $G_0$. There's only one person in the upper-class family so it's not really a consensus, though. The big leader listens to other people slightly and $c_j^* = b_{1j} + B_{1G_j} \eta_{G_j}$ – it's the summation of how the big leader listens to candidate $j$ and all the people who are supposed to support candidate $j$ in this family. It makes sense.

It's still more reasonable that

$$\hat{A}_{1,I+J} = \begin{bmatrix} \hat{a}_{11} & \hat{A}_{1G} & \hat{A}_{1J} \end{bmatrix}$$

In this case this big family with tree structure is an upper-class family. Again, the big leader collects opinions from people in the family by $\hat{A}_{1,I+J}$. $B_{1J}$ and $B_{1G}$ in Equation (5) are replaced by $\hat{A}_{1J}$ and $\hat{A}_{1G}$.

**Family with Equality**

A family doesn't need a leader. All the members are treated "equally" in a family like this:

$$\hat{A} = \begin{bmatrix} r & s & s \\ s & r & s \\ s & s & r \end{bmatrix}, r = 1 - 2s \geq 0, s > 0$$

For a family with $k$ members, $r = 1 - (k-1)s \geq 0$.

It's an upper-class family.

The downside of this family is that they won't almost support one candidate simultaneously. An army cannot be like this.

European Union is like this.

## VII. Conclusion

Politics is very complicated. The political status of a society definitely cannot be represented by a simple n-by-n matrix completely, just like statistics numbers cannot represent an athlete. However, some important and well-known properties of politics have been observed in this simple model already.

Politics is complicated even though the status is represented by a matrix only in this model. For example, when person $i$ decides to support a candidate, it can be his own decision, or it's a decision of person $j$ and person $i$ decides to listen to person $j$, or it's a decision of person $k$ when person $j$ decides to listen to person $k$ and person $i$ decides to listen to person $j$…

When all people in a society are assumed to be selfish, politics is unavoidable.

Anyway the wind blows.